\documentclass[nobibnotes,prb,showpacs,twocolumn]{revtex4}
\usepackage{amsmath,amssymb,bm,graphicx}

\newcommand{\bto}{BaTiO$_3$}
\newcommand{\magnetite}{Fe$_3$O$_4$}

\begin{document}

\title{Magnetic anisotropy modulation of magnetite
in \magnetite/\bto(100) epitaxial structures}

\author{C. A. F. Vaz}
\email[Corresponding author. Email: ]{carlos.vaz@yale.edu}
\author{J. Hoffman}
\author{A.-B. Posadas}
\author{C. H. Ahn}
\affiliation{Department of Applied Physics, Yale University, New
Haven, Connecticut 06520}%
\affiliation{Center for Research on Interface Structures and
Phenomena (CRISP), Yale University, New Haven, Connecticut 06520}%

\date{\today}

\begin{abstract}
Temperature dependent magnetometry and transport measurements on
epitaxial \magnetite\ films grown on \bto(100) single crystals by
molecular beam epitaxy show a series of discontinuities, that are
due to changes in the magnetic anisotropy induced by strain in the
different crystal phases of \bto. The magnetite film is under
tensile strain at room temperature, which is ascribed to the lattice
expansion of \bto\ at the cubic to tetragonal transition, indicating
that the magnetite film is relaxed at the growth temperature. From
the magnetization versus temperature curves, the variation in the
magnetic anisotropy is determined and compared with the
magnetoelastic anisotropies. These results demonstrate the
possibility of using the piezoelectric response of \bto\ to modulate
the magnetic anisotropy of magnetite films.
\end{abstract}



\maketitle

Complex oxides are characterized by a range of multifunctional
electronic behavior, including magnetism, ferroelectricity, and
correlated electron transport phenomena. Recently, much effort has
been devoted to controlling these properties by means of external
electric and magnetic fields. Besides intrinsic multiferroic
materials, which display simultaneous magnetic and electric
polarization,\cite{Fiebig05} new classes of composite materials are
being developed that explore the different properties of the
individual components to develop systems with larger
susceptibilities and more varied response functions. Examples of
multiferroic composites include \magnetite/BaTiO$_3$
heterostructures, where a strong bias dependence of the
magnetoresistance is observed.\cite{ZBPM06} Magnetite, \magnetite,
is a complex oxide material with several outstanding properties: it
is a ferrimagnetic conductor with a high Curie temperature of $\sim
850$ K and large spin polarization.\cite{Gorter55,VAH08} It is a
mixed valence oxide which crystallizes in the cubic inverse spinel
structure, with a room temperature lattice constant of 8.397 \AA.
Here, we exploit the fact that epitaxial \magnetite\ films grown on
lattice mismatched substrates develop an extra magnetic anisotropy
term due to magnetoelastic coupling. Of potential interest would be
the ability to modulate the magnetic anisotropy of magnetite by
using the piezoelectric effect of a ferroelectric material, such as
\bto. As a first step toward this goal, we consider the strain
induced by the natural phase transitions that occur in \bto\ as a
function of temperature in order to study the changes in the
magnetic anisotropy of a thin \magnetite\ film. This approach has
been employed by other authors to study magnetoresistance effects in
La$_{1-x}$Sr$_x$MnO$_3$ films\cite{LNE+00,DFBS03} and the magnetic
anisotropy of CoFe$_2$O$_4$\cite{CS06} and Fe.\cite{SPD+07} Very
recently, Tian et al.\cite{TQL+08} have reported strain-induced
changes in the magnetization of \magnetite/\bto, which were
attributed to changes in the magnetic domain structure of
\magnetite. In these films the authors suggest the \magnetite\ is
compressively strained. In this letter, we show that the magnetic
anisotropy of thin \magnetite\ films is modulated by the change in
strain associated with the structural phase transitions of \bto. We
demonstrate that at room temperature, the magnetite films are under
tensile strain, which is key to understanding the changes in the
magnetic anisotropies. We explain the changes in magnetic anisotropy
in terms of magnetoelastic coupling and provide numerical estimates
for the various \bto\ phases and domains.

\bto\ has a simple cubic perovskite structure between 1733 K and 393
K ($a = 3.996$ \AA\ at 393 K). Between 393 K and 278 K, it is
tetragonal and ferroelectric, with lattice constants $a=3.9920$ \AA\
and $c=4.0361$ \AA\ (293 K). Below 278 K it converts to an
orthorhombic structure ($a = b = 4.013$ \AA, $c = 3.99$ \AA,
$\gamma=89.869^\mathrm{o}$, in the monoclinic representation), and
below 183 K it undergoes a further transition to a rhombohedral
structure ($a = 4.001$ \AA, $\gamma=89.85^\mathrm{o}$ at $T=105$
K).\cite{AAA+02} In all cases, the lattice can be seen as a slightly
distorted cubic structure.

\begin{figure}[t!]
\centering
\includegraphics[width=8.5cm]{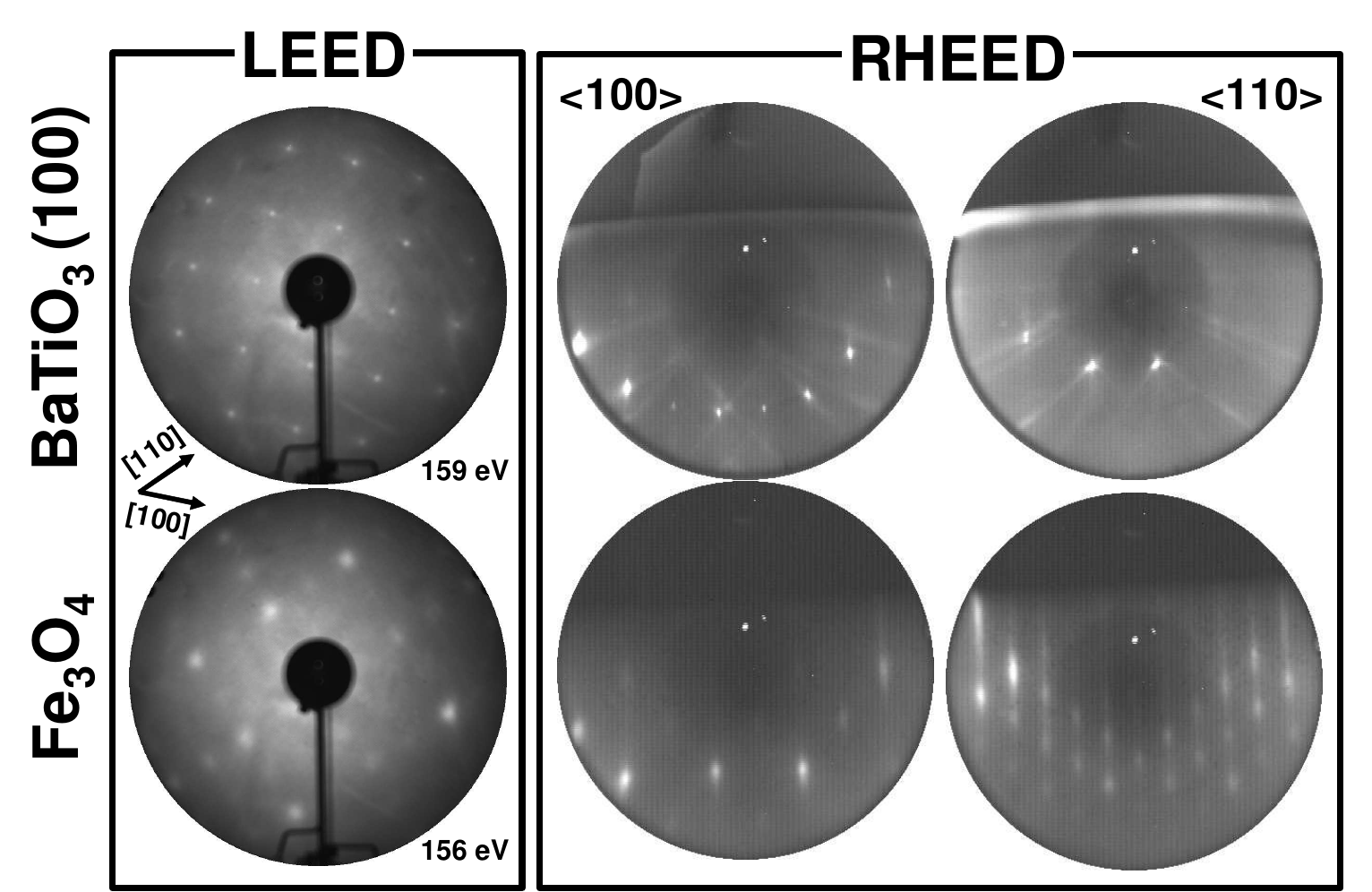}
\caption{LEED and RHEED patterns of the \bto\ substrate and the
\magnetite\ film at room temperature. The incident electron beam
energy for RHEED is 15 keV, and the RHEED azimuths are with respect
to the incident beam direction.} \label{fig:EED}
\end{figure}

For this study, a 30 nm thick magnetite film was grown on a
\bto(100) single crystal by molecular beam epitaxy in an ultrahigh
vacuum deposition system with a base pressure of $1\times 10^{-9}$
mbar. Prior to film growth the \bto\ substrate was annealed at 900
K, exhibiting sharp low energy electron diffraction (LEED) and
reflection high energy electron diffraction (RHEED) patterns, as
shown in Fig.~\ref{fig:EED}. For the magnetite growth, an atomic Fe
beam was thermally generated from an effusion cell under an O$_2$
partial pressure of $2\times 10^{-7}$ mbar, with the substrate
temperature being held at 570 K. The film thickness was estimated
from a calibrated thickness monitor (and corroborated by x-ray
reflectivity measurements), while film crystallinity was monitored
during growth using RHEED. After growth, the film was characterized
{\it in situ} by LEED and x-ray photoemission spectroscopy (XPS).
{\it Ex situ} measurements include x-ray diffraction (XRD), atomic
force microscopy (AFM), superconducting quantum interference device
(SQUID) magnetometry, and resistivity measurements; the \bto\
substrate crystal orientation was confirmed by Laue diffraction.

\begin{figure}[tbh]
\centering
\includegraphics[width=8.0cm]{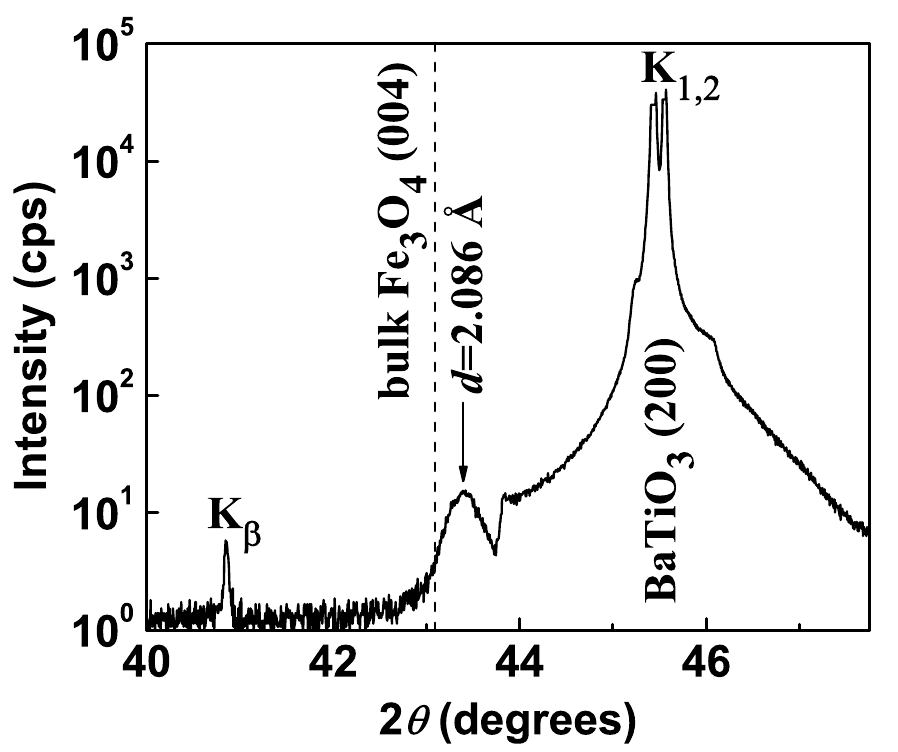}
\caption{High resolution $\theta-2\theta$ XRD measurements (Cu
$K_\alpha$) near the (002) scattering plane of \bto.}
\label{fig:XRD}
\end{figure}

RHEED and LEED patterns indicate that magnetite grows epitaxially on
\bto\ in the (100) orientation, with
\magnetite(100)[100]$\parallel$\bto(100)[100].\cite{TQL+08} The
RHEED diffraction spots of \magnetite\ broaden compared to those of
the \bto\ during the early stages of growth, indicating a
three-dimensional growth mode. The observation of LEED patterns in
the magnetite film indicates a relatively well ordered surface. XPS
measurements show the presence of both Fe$^{2+}$ and Fe$^{3+}$, as
expected for the mixed valence magnetite. These data rule out the
growth of the competing maghemite phase ($\gamma$-Fe$_2$O$_3$, also
a cubic inverse spinel with $a = 8.322$ \AA) and indicate that we
are able to grow stoichiometric, high quality epitaxial \magnetite\
thin films on \bto. AFM shows that the \magnetite\ surface is
relatively smooth, with an average surface roughness of 2 \AA\ over
4 $\mu$m$^2$ scan areas. XRD measurements after sample growth
indicate that, in this case, the \bto\ contains no $c$ domains (see
Fig.~\ref{fig:XRD}), and no planes other than the (100) are
detected. The peak at $\theta = 43.39^\mathrm{o}$ is attributed to
\magnetite, giving an out of plane lattice constant of 8.344 \AA,
smaller than the bulk value (8.397 \AA) and at odds with what is
expected from the lattice mismatch with \bto(100) of about 4.7\%.
This result can be explained by assuming that the magnetite film
relaxes fully during growth, while the tensile strain results from
the lattice expansion of \bto\ at the cubic to tetragonal transition
at 393 K.

\begin{figure}[tbh]
\centering
\includegraphics[width=8.5cm]{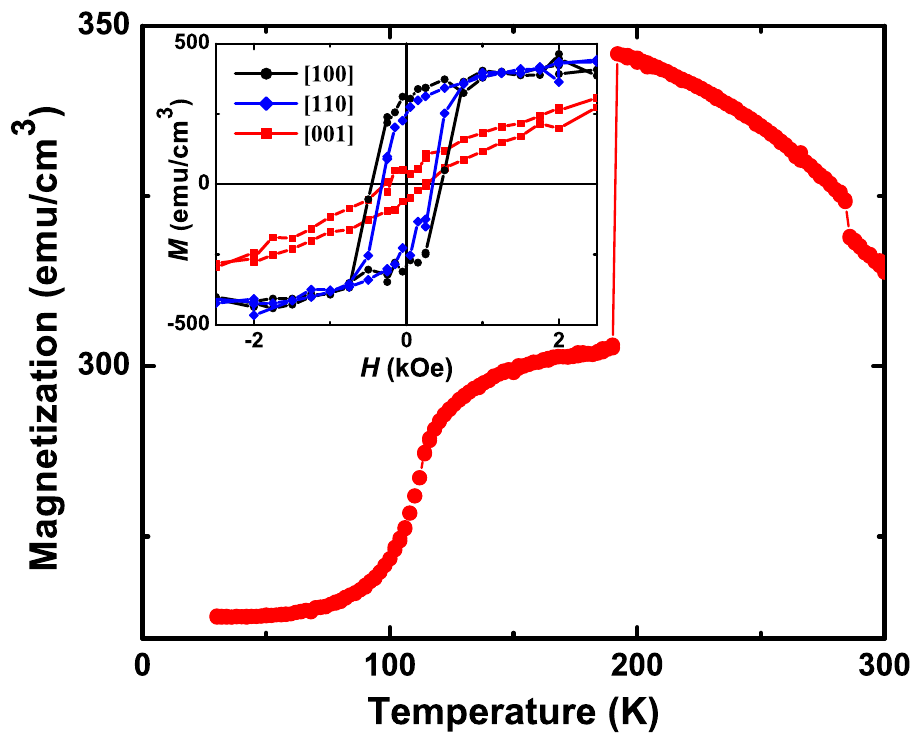}
\caption{Magnetization versus temperature dependence of magnetite
along the in-plane [100] direction under no applied field. The
\magnetite\ film was initially saturated magnetically at 320 K under
a field of 10 kOe. The inset shows room temperature {\it M-H} loops
along different crystal directions, corrected for the diamagnetic
contribution from the substrate. The [001] direction is out of
plane.} \label{fig:MT_0Oe}
\end{figure}

Figure~\ref{fig:MT_0Oe} shows the variation of the magnetic moment
of magnetite as a function of temperature at zero applied field. We
find that at the temperatures corresponding to the phase transitions
of BaTiO$_3$, there are large jumps in the Fe$_3$O$_4$
magnetization, which we associate with changes in the domain
structure due to strain-induced variations in the magnetic
anisotropy. This result demonstrates that we are able to modulate
the magnetic response of magnetite via a strain-induced
magnetoelastic interaction. Note in particular the large change in
magnetization at around 120 K, which is coincident with the
vanishing of the magnetocrystalline anisotropy constant at the
Verwey temperature of bulk magnetite.\cite{Lefever70}  {\it M-H}
curves at room temperature show that the film is magnetized
in-plane, although there is a small non-zero remanence along the out
of plane direction, as shown in the inset to Fig.~\ref{fig:MT_0Oe}.
From our data we obtain a saturation magnetization $M_s = 470 \pm
40$ emu/cm$^3$, similar to the bulk value. Resistivity vs
temperature measurements (not shown) also reveal the presence of
discontinuities at the \bto\ phase transitions, which we associate
with changes in domain structure and to anisotropic
magnetoresistance; a large increase in resistivity by several orders
of magnitude at about 120 K compares well with the onset of the
Verwey transition, indicating a magnetite film with good
stoichiometry.

\begin{figure}[b!]
\centering
\includegraphics[width=8.5cm]{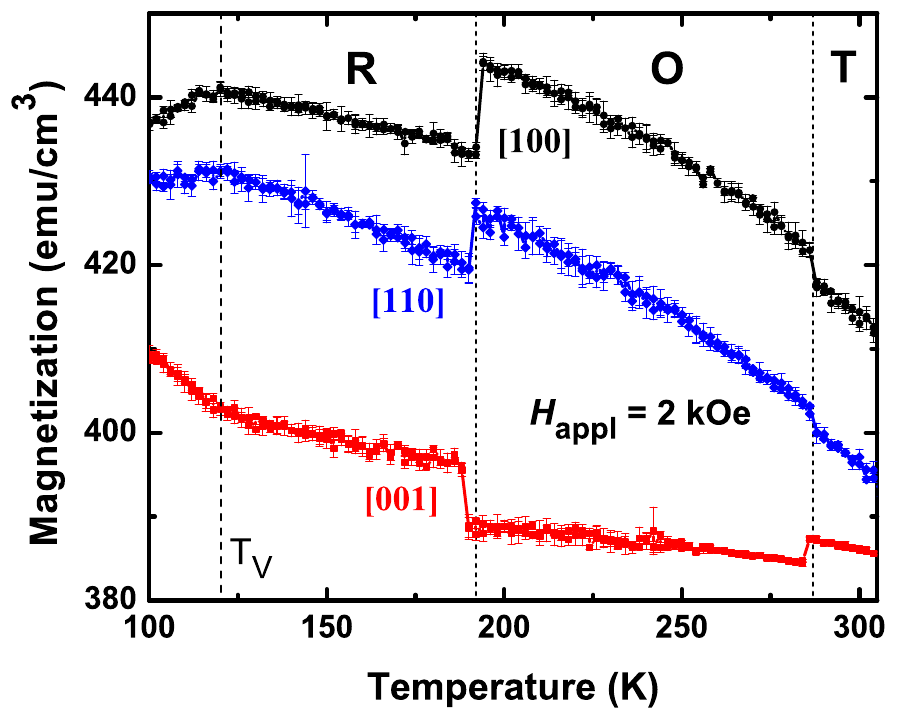}
\caption{Temperature variation of the Fe$_3$O$_4$ magnetization
under an applied field of 2 kOe, along different crystallographic
directions, as labeled. Measurements were performed during cooling
in the sequence [100], [110] and [001]; the latter corresponds to
the out of plane direction. Curves have been shifted vertically for
convenient data display and substrate diamagnetic contributions have
not been removed.} \label{fig:MT_2kOe}
\end{figure}

To study the changes in the magnetic anisotropy more systematically,
we measured the variation of the magnetic moment of magnetite as a
function of temperature along different crystal directions under a
constant magnetic field of 2 kOe, which lies in the reversible part
of the {\it M-H} curve but below saturation. In this case, changes
in the magnetization reflect directly changes in the magnetic
anisotropy rather than in the domain structure. The results of the
magnetization measurements are shown in Fig.~\ref{fig:MT_2kOe},
where again we find that at the temperatures corresponding to the
phase transitions of BaTiO$_3$, there are abrupt jumps in the
magnetization. In particular, the jumps in the two in-plane
magnetization curves are in the same direction, increasing at 287 K
and decreasing at 192 K, while the out of plane magnetization curve
has the opposite behavior.\cite{TQL+08} This clearly indicates that
the changes in anisotropy occur predominantly as a consequence of
changes in the out of plane magnetic anisotropy, which is reduced
between 287 K and 192 K. We attribute the variation of the magnetic
anisotropy observed experimentally at the phase transitions of \bto\
to strain-induced magnetoelastic anisotropy, arising from
distortions in the crystal structure of the magnetite film.

The magnetic anisotropy can be estimated from the phenomenological
expressions for the elastic and magnetoelastic energies,
yielding:\cite{Vaz08}
\begin{equation}
\nonumber %
E_\mathrm{me}= [K_\mathrm{t}\cos 2\phi + K_\mathrm{o} \sin 2\phi
]\sin^2\theta + K_\mathrm{p}\cos^2\theta
\end{equation}
where $K_\mathrm{t}= b_1(\epsilon_{xx}-\epsilon_{yy})/2$ and
$K_\mathrm{o}=b_2\epsilon_{xy}/2$ correspond to in-plane
magnetoelastic anisotropy constants, and
$K_\mathrm{p}=b_1(1+c_{11}/2c_{12})\epsilon_{zz}$ is the
perpendicular magnetoelastic anisotropy constant. In this
expression, $\epsilon_{ij}$ are the strain components, $c_{ij}$ the
elastic constants, $b_i$ the magnetoelastic coupling
coefficients,\cite{Chikazumi97} $\theta$ and $\phi$ the polar and
azimuthal angles of the spherical coordinate system, respectively,
with the polar axis along the $z$ direction. For tensile strain, the
magnetoelastic interaction in \magnetite\ favors perpendicular
magnetic anisotropy ($K_\mathrm{p}<0$), in agreement with our
observation of a negative out of plane strain in XRD and the
presence of perpendicular magnetic remanence in the {\it M-H} loops.
This result is not expected if a positive (compressive) residual
strain were present from the lattice mismatch between \magnetite\
and \bto. In Table~\ref{tab:Kp} we show our estimates for
$K_\mathrm{p}$, assuming the experimental strain at 300 K and
considering the strain that results from the lattice distortions of
the \bto\ (we take into account the temperature variation of both
\bto\ and \magnetite\ lattice constants). Negative $K_\mathrm{p}$
favors perpendicular magnetization; these simple calculations
suggest that $K_\mathrm{p}$ can change by a factor of 4 from the
square surface domains of tetragonal \bto\ to the oblique surface
domains of the orthorhombic phase.

\begin{table}[tbh]
\caption{Calculated values for $10^6K_\mathrm{p}$ (erg/cm$^3$), for
\magnetite\ films strained according to the different phases and
domains of the underlying \bto\ substrate at 272 K and 187 K.}
\label{tab:Kp}
\begin{ruledtabular}
\begin{tabular}{cccccccc}
$T$(K) & \multicolumn{3}{c}{tetragonal} & \multicolumn{3}{c}{orthorhombic} & rhombohedral \\
\cline{2-3} \cline{5-6} \cline{8-8}
& square & rectangle && rectangle & oblique && oblique\\
\hline %
272 & -0.22 & -0.71 && -0.47 & -0.75 && ---\\
187 & --- & --- && -0.45 & -0.85 && -0.66\\
\end{tabular}
\end{ruledtabular}
\end{table}

Thermal cycling of the \bto\ crystal  during the temperature
dependent measurements leads to the formation of different types of
domains, for instance, $a$ and $c$ domains in the tetragonal phase,
and oblique or rectangular in the orthorhombic phase. These domains
give rise to different strains in the magnetite film and to
different magnetoelastic anisotropy contributions to the magnetic
energy. In the tetragonal phase, rectangular domains result in large
perpendicular magnetic anisotropy; in the orthorhombic phase,
$K_\mathrm{p}$ is strongly reduced for rectangular domains, while it
increases slightly in oblique domains. Therefore, the decrease in
the perpendicular anisotropy can be explained by a transformation of
$c$ domains of the tetragonal phase to similar domains in the
orthorhombic phase. At 192 K, the increase in perpendicular magnetic
anisotropy is explained by the prevalence of rectangular domains in
the orthorhombic phase. It suggests that our \bto\ crystal changes
from an $a$, $c$ multi-domain state at 300 K, to rectangular below
287 K and oblique below 192 K. We note that the magnetostatic energy
(shape anisotropy, which favors in-plane magnetization) of the
\magnetite\ film is of the order of $2\pi M_s^2 = 1.39 \times 10^6$
erg/cm$^3$. This value implies that the Fe$_3$O$_4$/\bto\ films
would remain magnetized in-plane, assuming a uniform strain state;
in real films, one may expect large interface strains and larger
magnetoelastic anisotropies, which may be at the origin of the
observed small perpendicular remanent magnetization.

In conclusion, we show that we are able to grow high quality
\magnetite/\bto\ heterostructures by molecular beam epitaxy. A
tensile strain is present in the \magnetite\ film at room
temperature, which indicates that the film relaxes at the growth
temperature due to the large lattice mismatch. We show, based on
calculations of the magnetoelastic anisotropy constants, that such
tensile strain gives rise to a strong perpendicular magnetic
anisotropy, as observed experimentally. The different lattice
distortions induced by the phase transition of \bto\ lead to strong
modifications in the magnetocrystalline anisotropy, manifested by
large discontinuities of zero-field and near saturation {\it M-T}
curves, demonstrating magnetic anisotropy modulation of \magnetite\
films via strain coupling to the \bto\ substrate. These types of
structures may be used to control magnetoelectric devices, such as
the on/off state of planar Hall effect devices,\cite{BKY+06} or to
study the effect of strain on the Verwey transition of magnetite.

The authors acknowledge financial support by the NSF through MRSEC
DMR 0520495 (CRISP), NRI, ONR, and the Sloan and Packard
Foundations.


\end{document}